# Analysis of a Memcapacitor-Based for Neural Network Accelerator Framework

*Ankur Singh[1], Dowon Kim[1], and Byung-Geun Lee[1]\**

[1]School of Electrical Engineering and Computer Science, Gwangju Institute of Science and Technology, Gwangju 61005, Republic of Korea (e-mail: ankursingh@gm.gist.ac.kr; kimdowon1109@gist.ac.kr; bglee@gist.ac.kr).

**Abstract:** Data-intensive computing tasks, such as training neural networks, are crucial for artificial intelligence applications but often come with high energy demands. One promising solution is to develop specialized hardware that directly maps neural networks, utilizing arrays of memristive devices to perform parallel multiply–accumulate operations. In our research, we introduce a novel CMOS-based memcapacitor circuit validated using the cadence tool. Additionally, we developed the device in Python to facilitate the design of a memcapacitive-based accelerator. Our proposed framework employs a crossbar array of memcapacitor devices to train a neural network capable of digit classification and CIFAR dataset recognition. We tested the non-ideal characteristics of the constructed memcapacitor-based neural network. The system achieved an impressive 98.4% training accuracy in digit recognition and 94.4% training accuracy in CIFAR recognition, highlighting its effectiveness. This study demonstrates the potential of memcapacitor-based neural network systems in handling classification tasks and sets the stage for further advancements in neuromorphic computing.

## 1. Introduction

Memelements have emerged as a promising class of devices, demonstrating remarkable performance, particularly when deployed in crossbar architectures [1-3]. Their integration into these structures significantly enhances the efficiency of vector-matrix multiplication (VMM) by enabling the parallel execution of product and summation operations through the devices. This capability is particularly beneficial in the domain of convolutional neural networks (CNNs), where extensive matrix operations are fundamental to both training and inference processes. The combination of in-memory computing (IMC) architectures with the adjustable analog memductance of memelements further contributes to power-efficient VMM and training, enabling the development of highly integrated memory architectures. Consequently, a wide array of CNN hardware designs utilizing memelements-based VMM accelerators [3-6] has been proposed, with their effectiveness consistently demonstrated in various studies.

Neuromorphic computing, modeled after brain-like processes and grounded in artificial neural networks, presents effective solutions for a wide range of computationally demanding tasks. Originally conceptualized in the 1980s [7-8], this field has seen substantial progress with the advent of memristive devices [9] and the introduction of convolutional layers in deep neural networks [10-11]. These innovations have facilitated the development of various resistive neuromorphic systems that employ materials such as oxides [12-14], phase-change memory [15], spintronic devices [16-17], and ferroelectric components, including ferroelectric tunnel junctions [18-19] and ferroelectric field-effect transistors (FeFETs) [20-21]. Among these, devices like ferroelectric tunnel junctions and silicon–oxide–nitride–oxide–silicon (SONOS) transistors have achieved remarkable energy efficiencies, reaching up to 100 tera-operations per second per watt (TOPS/W) [22]. These neuromorphic systems operate by storing synaptic weights in analog form for multiplication tasks, utilizing Kirchhoff's current law to sum currents through crossbar arrays.

Memcapacitive devices, akin to memristive devices but operating on a capacitive principle, hold the promise of lower static power consumption [23]. While theoretical models for memcapacitive devices have been extensively explored [23-27], practical implementations remain relatively scarce [28-31]. These devices can be realized through various mechanisms, such as implementing a variable plate distance concept in micro-electromechanical systems [32], utilizing a metal-to-insulator transition material in series with a dielectric layer [27], modulating the oxygen vacancy front in a traditional memristor [25], and employing a simple metal–oxide–semiconductor capacitor with memory effects [29, 30]. Despite the potential advantages, achieving a high dynamic range in memcapacitive devices poses significant challenges. For instance, devices with a variable plate distance often encounter large parasitic resistive components when the plate distance is minimal, hindering their performance [25]. Conversely, at large plate distances, these devices face limited lateral scalability. Similar issues arise in memcapacitors designed with varying surface areas [32] or dielectric constants [31], as these approaches also struggle with scalability and efficiency.

In our study, we introduce a cutting-edge memcapacitor-based neural network framework tailored for data-intensive computing tasks, such as neural network training, which are essential for artificial intelligence applications but often involve significant energy consumption. We present a novel CMOS-based memcapacitor circuit, validated using the cadence tool, and developed a Python-based model to facilitate the design of a memcapacitive-based accelerator. This framework employs a 5 x 5 x 5 crossbar array of memcapacitor devices to train a neural network for digit classification and CIFAR dataset recognition. Our system effectively addresses the non-ideal characteristics of memcapacitive devices, achieving remarkable training accuracies of 98.4% in digit recognition and 94.4% in CIFAR recognition. This research demonstrates the efficacy of memcapacitor-based neural networks in handling classification tasks, showcasing their potential in reducing the energy demands of data-intensive computing. The integration of memcapacitor technology within neural networks not only enhances computational efficiency but also underscores the viability of these systems in neuromorphic computing. Our work sets the stage for further advancements by illustrating the practical applications of memcapacitors in neural network training and classification tasks. The significant training accuracies achieved highlight the promise of memcapacitor-based systems in AI, suggesting a substantial impact on future developments in energy-efficient and high-performance computing.

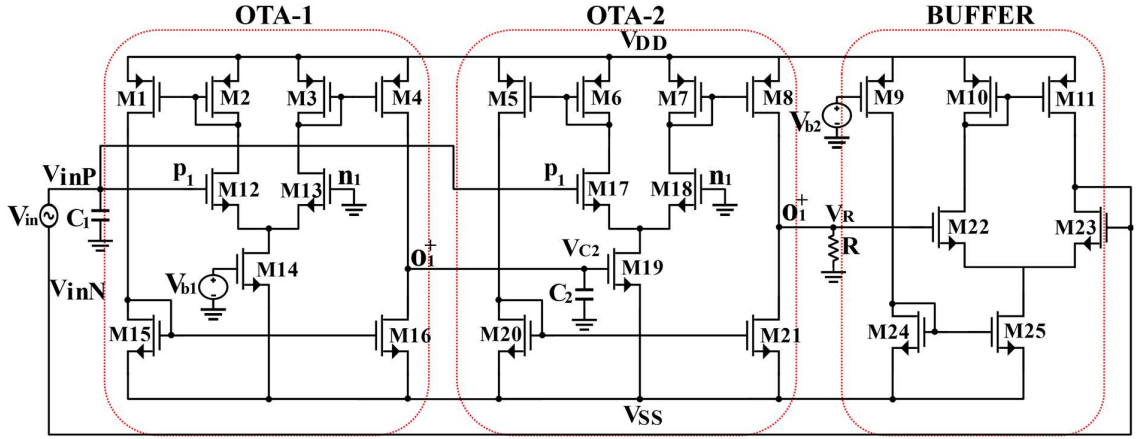

Fig. 1. Proposed MOSFET based memcapacitor circuit.

## 2. Proposed Memcapacitor Device

### 2.1 CMOS based model

The memcapacitor represents the relationship between the time integral of charge and flux, defined by the inverse memcapacitance function $M_C(q)$, with constants α and β. The mathematical model of a charge-controlled memcapacitor is expressed as follows [33].

$$M_C = \frac{V_{in}(t)}{q(t)} = \beta \pm \alpha \sigma(t) \tag{1}$$

The emulator proposed in this work comprises three functional blocks: two operational transconductance amplifiers (OTA) and a buffer. These blocks are interconnected to achieve the desired memcapacitor characteristic, which relates the input voltage Vin to the charge q(t). The overall design of the memcapacitor emulator is illustrated in Fig. 1 [34]. In this design, the body terminals of all PMOS and NMOS transistors are connected to $V_{DD}$ and $V_{SS}$, respectively. The port characteristics of the OTA [35] can be mathematically defined by the following equations:

$$I_{O_i^+} = g_{mi}(V_{pi} - V_{ni}) \tag{2}$$

where $g_{mi}$, $V_{pi}$, and $V_{ni}$ represents the transconductance gain and input voltage of each OTA. The routine analysis yields the following expression for gm:

$$g_{mi} = K(V_{bi} - V_{SS} - V_{th}) \tag{3}$$

Where "i" is the OTA number, Vss is the supply voltage of the OTA, Vth is the threshold voltage of MOSFET device and "K" is a parameter of the MOSFET device given by

$$K = \mu_n C_{OX} \frac{W}{L}$$

In this formulation, W represents the channel width, L denotes the channel length, $\mu_n$ is the mobility of the carrier, and $C_{OX}$ refers to the oxide capacitance per unit area in the MOSFET.

The detailed working principle of the proposed emulator is analyzed through these functional blocks. OTA-1 and OTA-2, both OTAs, are crucial in defining the dynamic relationship between the charge and the flux. OTA-3, a buffer, ensures proper signal conditioning and stability of the emulator output. This configuration effectively replicates the theoretical behavior of a memcapacitor, providing a practical and implementable circuit design. The analysis shows how the integration of these blocks results in the desired emulation between the input voltage and the resulting charge over time.

In OTA-1, the positive terminal of the input signal VinP is connected. It is important to note that Vin's positive and negative terminals are labelled VinP and VinN, respectively. The transistors M12 and M13 form an input differential pair and are biased by the current sink transistor M14, as illustrated in Fig. 1 for OTA-1. This functional block converts the differential voltage into a current, as described below.

$$I_{O_1^+} = g_{m1}(V_{inP} - V_{n1}) \tag{4}$$
$$I_{O_1^+} = g_{m1}V_{inP}$$

In this configuration, $g_{m1}$ represents the transconductance of OTA-1, with $V_{n1}$ grounded as depicted in Figure 1. Functional Block-3 functions as a buffer, where the input signal is applied to the gate of transistor M23. Meanwhile, the gate of transistor M22 is connected to its drain terminal, establishing a negative feedback loop as illustrated in Figure 1. The corresponding analytical model for Functional Block-3 is provided below.

$$I_{22} = g_{22}(V_{P'} - V_X) = g_{23}(V_{N'} - V_X) \tag{5}$$

Given that $g_{22}=g_{23}$, it follows that $V_{P'}=V_{N'}$. Consequently, $V_{N'}$ at the output terminal mirrors the input $V_{P'}$, which is essential for ensuring the reliable operation of the proposed emulator.

The behaviour of the memcapacitor is demonstrated through the analysis of the functional blocks shown in Fig. 1. A sinusoidal input signal is applied to the proposed emulator between $V_{inP}$ and $V_{inN}$, as illustrated below.

$$V_{in} = V_{inP} - V_{inN} \tag{6}$$

In the proposed design, a sinusoidal signal is applied to $V_{in}$ across $C_1$, causing the input current ($I_{in}$) to flow through it. The voltage across $C_1$ is mathematically expressed as follows.

$$V_{C_1} = V_{inP} = \frac{1}{C_1}\int I_{in} dt = \frac{q(t)}{C_1} \tag{7}$$

The output current of OTA-1 is $(g_{m1}*q(t))/C1$. The output terminal of OTA is connected to $V_{b2}$ of OTA-2, which is the voltage across $C_2$ and can be expressed as,

$$V_{C_2} = V_{b2} = \frac{1}{C_2}\int g_{m1}V_{inP}dt = \frac{g_{m1}}{C_1C_2}\sigma(t) \quad (8)$$

Changing $C_2$ alters $V_{b2}$ due to the variation in charge, where $V_{b2}$ can be referred to as the charge control voltage. Upon analyzing OTA-2, the output current can be expressed by the following equation.

$$\begin{aligned} I_{O_2^+} &= g_{m2}(V_{p2} - V_{n2}) \\ I_{O_2^+} &= g_{m2}V_{inP} \end{aligned} \quad (9)$$

The above current flows through R generating a voltage $V_R$ as shown below:

$$\begin{aligned} V_R &= I_{O_2^+}R \\ V_R &= g_{m2}V_{inP}R \end{aligned} \quad (10)$$

Substituting $V_{b2}$ depicted in eq. 8 in eq. 3, $g_{m2}$ can be expressed as:

$$g_{m2} = K\left[\frac{g_{m1}\sigma(t)}{C_1C_2} - V_{SS} - V_{th}\right] \quad (11)$$

Now putting eq. 11 into eq. 10 and $V_R = V_{P'} = V_N$ from the buffer, we get

$$V_R = V_{P'} = V_{inN} = \frac{q(t)RK}{C_1}\left[\frac{g_{m1}\sigma(t)}{C_1C_2} - V_{SS} - V_{th}\right] \quad (12)$$

Consequently, by Substituting eq. 12 into 6, the corresponding memcapacitance can be calculated as follows

$$\begin{aligned} V_{in}(t) &= \frac{q(t)}{C_1} - \frac{q(t)RK}{C_1}\left[\frac{g_{m1}\sigma(t)}{C_1C_2} - V_{SS} - V_{th}\right] \\ M_C &= \frac{V_{in}(t)}{q(t)} = \frac{1}{C_1}[1+V_{SS}+V_{th}] - \frac{RK}{C_1}\left[\frac{g_{m1}\sigma(t)}{C_1C_2}\right] \end{aligned} \quad (13)$$

The proposed design features a meminductor model, initiating with a memcapacitance value calculated as $[(1/C1)(1+V_{SS}+V_{th})]$ according to the formula provided above. The rate at which the capacitance changes is determined by $[((RK/C1)(g_{m1}*\sigma(t))/C_1*C_2)]$. Here gm1 represent the transconductance values of the OTA-1 [36].

The simulation setup for the proposed work comprises Cadence Virtuoso software on the TSMC 180-nmtechnology node. All the simulations were carried out in Analog Design Environment (ADE) window having 27°C as the nominal temperature setting and the plot shown in Fig. 2 [37]. Also, the proposed work employs only a positive power supply i.e., $V_{DD}$ which is taken to be $1.8V_T$.

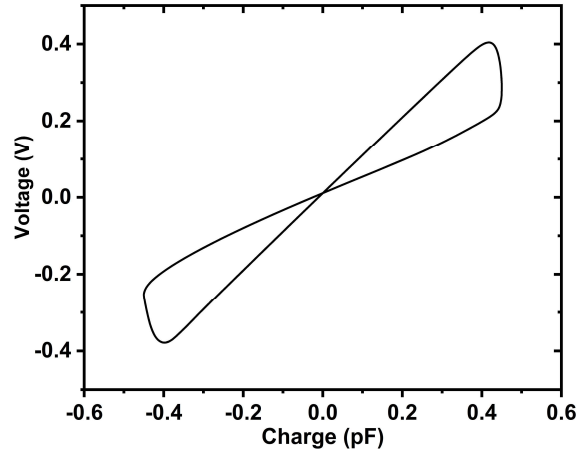

Fig. 2. Hysteresis curve of the proposed memcapacitor circuit.

## 2.1 Memcapacitor spice model

The proposed memcapacitor circuit is meticulously designed using TSMC 180nm CMOS technology, tailored for implementing vector-matrix multiplication—a fundamental operation in CNN. This circuit leverages the non-linear characteristics of memcapacitors to perform computational tasks typically associated with neuromorphic computing. The model is built using a combination of NMOS and PMOS transistors with varying channel lengths and widths, carefully selected to optimize the circuit's performance in terms of speed, power consumption, and accuracy. The circuit architecture involves the strategic placement of MOSFETs (Metal-Oxide-Semiconductor Field-Effect Transistors) that act as switches, enabling the memcapacitor to store and manipulate charge in a way that emulates synaptic behavior in biological neurons. The SPICE model incorporates three different CMOS configurations to accommodate the various transistor dimensions required for different stages of the circuit [38]. Additionally, passive components like capacitors and resistors are integrated into the design to stabilize the circuit and control the flow of current.

A sinusoidal voltage source is applied as the input to the circuit, with parameters such as amplitude and frequency finely tuned to mimic the dynamic signals encountered in neuromorphic systems [39]. The transient simulation of this circuit is performed using a step-by-step approach, where the time-dependent voltage across different nodes is analyzed to verify the memcapacitor's performance. This model is pivotal for simulating the behavior of the memcapacitor in a controlled environment, providing insights into its potential application in energy-efficient, high-performance neuromorphic computing [40]. To systematically develop and validate the memcapacitor model, an algorithm has been structured, as illustrated in Fig. 1. The algorithm outlines the following critical steps for implementing the memcapacitor model in Python.

---

**Algorithm 1 Construct Memcapacitor model**

    ***Input:***
    $C_{name}$: *Define circuit name*
    $L_{CMOS}$: *CMOS libraries*
    $M_{NMOS}$, $M_{PMOS}$: *Transistor models*
    $C_1$, $C_2$: *Capacitors*
    $R_1$: *Resistor*
    $V_{DD}$, $V_{SS}$, $V_{bias}$: *Voltage sources*
    $V_{in}$: *Input signal*
    $P_{sim}$: *Simulation parameters*
    ***Output:*** *Simulated transient response, voltage plots, circuit performance metrics, capacitance*
    ***Initialize*** $C_{name}$ *and include* $L_{CMOS}$
    ***Define*** *transistor parameters* $M_{NMOS}$, $M_{PMOS}$ *(e.g.* ***W, L****)*
    ***Construct the Circuit:***
        *Add transistor* $M_{NMOS}$, $M_{PMOS}$
        *Insert Capacitors* $C_1$, $C_2$
        *Place resistor* $R_1$
    ***Apply*** *voltage sources* $V_{DD}$, $V_{SS}$, $V_{bias}$
    ***Configure*** *input signal* $V_{in}$
    ***Run*** *transient simulation with* $P_{sim}$
    ***Analyze and plot simulation results***

---

## 3. Memcapacitive-based Vector Matrix Multiplication

The Memcapacitor-based VMM is a novel and promising approach in the field of neuromorphic computing, where the goal is to mimic the way the human brain processes information [41]. In traditional digital systems, VMM is a core operation in machine learning algorithms, particularly in CNN. However, as these systems scale, the energy consumption and speed of conventional hardware become significant bottlenecks [42]. This is where memcapacitors, with their unique ability to retain information and perform non-linear operations, offer a distinct advantage. Memcapacitors are passive electronic components that combine the properties of memory and capacitance. Unlike traditional capacitors, which store charge linearly, memcapacitors can store different amounts of charge based on their history of applied voltage, making them ideal for use in systems that require complex, non-linear operations such as VMM. In a memcapacitive VMM circuit, the stored charge in each memcapacitor represents the weight of a connection in a neural network, while the input voltage corresponds to the input signals of the neurons [43]. This setup allows the circuit to perform multiplication and accumulation in a single step, significantly reducing the computational overhead. The VMM can be represented mathematically as:

$$(y_1....y_n) = (x_1....x_m) \cdot \begin{pmatrix} w_{11} & \cdots & w_{1n} \\ \vdots & \ddots & \vdots \\ w_{m1} & \cdots & w_{mn} \end{pmatrix}$$

The output vector y is calculated by the inner product of the input vector x and weight matrix w and each element of the output vector yi can be expressed as follows:

$$y_i = \sum_{k=1}^{m} x_k \cdot w_{ki}$$

In memcapacitor-based VMM, as shown in Fig. 3, the weight matrix is implemented by a crossbar memcapacitor array with m-rows and n-columns [44]. The rows and columns of the memcapacitor array are connected to m input pulses and n output, respectively.

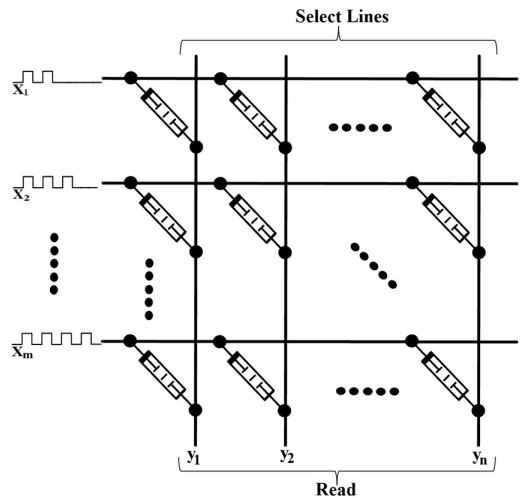

Fig. 3. Crossbar structure defined in the simulator using memcapacitor device.

## 4. Results and Discussion

In this section, we utilized both the MNIST and CIFAR datasets for training and inference within our system. The MNIST dataset, a widely recognized benchmark in image processing, consists of 70,000 grayscale images of handwritten digits, each measuring 28 x 28 pixels and representing digits from 0 to 9. For our experiments, we divided this dataset into 40,000 training images and 20,000 testing images [45]. The CIFAR dataset, another prominent dataset in the field of machine learning, comprises 60,000 color images divided into 10 classes, with 50,000 images designated for training and 10,000 for testing [46]. The core of our work involves the meticulous design of a memcapacitor circuit using TSMC 180nm CMOS technology, as outlined in Algorithm 1, with the primary objective of implementing VMM for image classification tasks. The memcapacitor's ability to store different charge levels, which represent the weights in a neural network, is central to this VMM operation. This circuit design is emulated in Python, where we applied the memcapacitor-based model to implement a CNN for digit recognition using the MNIST dataset. The combination of MNIST and CIFAR datasets allowed us to validate the versatility and effectiveness of the proposed system in different image classification scenarios [47].

In our proposed neural network architecture, we implemented a 5 x 5 x 5 vector matrix multiplication (VMM) utilizing a memcapacitor SPICE model as the first convolutional layer. To optimize computational efficiency, the MNIST dataset was resized from its original 28 x 28 dimensions to 20 x 20 pixels before being processed through the memcapacitor-based VMM. As depicted in Fig. 4, this VMM leverages the unique properties of memcapacitors to perform computations, with initial capacitance values determined by Equation 13. During each training epoch, the VMM weights representing capacitance values were iteratively adjusted to enhance model performance. The processed output from the VMM is subsequently passed through a neural network that includes layers with ReLU activation functions and a dense layer with 64 units, culminating in a softmax output layer for image classification. This method harnesses the computational efficiency of memcapacitor-based VMMs while integrating the adaptability of neural networks, resulting in an effective and accurate image classification framework. The training accuracy of 98.4% and a testing accuracy of 94.6% as shown in Fig. 5. The algorithm 2 is used implement the simulator for classification of MNIST dataset. Additionally, Fig. 6 provides a visual representation of the confusion matrix, illustrating the correlation between the true labels and the predicted labels, thereby offering insights into the model's performance. Fig. 7 showcases the optimized weights within the memristor crossbar array. These weights are indicative of the learned patterns that enable the model to perform with high accuracy. We further extended our approach to the CIFAR dataset for both training and inference. In this scenario, two convolutional layers were implemented using memcapacitor-based VMM. The first convolutional layer utilized a 5 x 5 x 5 VMM, while the second layer employed a 5 x 5 x 15 VMM configuration, as depicted in Fig. 8. These VMM, which leverage the distinct characteristics of memcapacitors, efficiently processed the input data. The output from these convolutional layers was subsequently fed into a neural network architecture consisting of layers with ReLU activation functions. This was followed by a dense layer with 128 units, culminating in a softmax output layer for image classification [48-49]. This approach yielded a training accuracy of 94.4% and a testing accuracy of 89.8% as shown in Fig. 9. The algorithm used for this process is analogous to the one employed for the MNIST dataset.

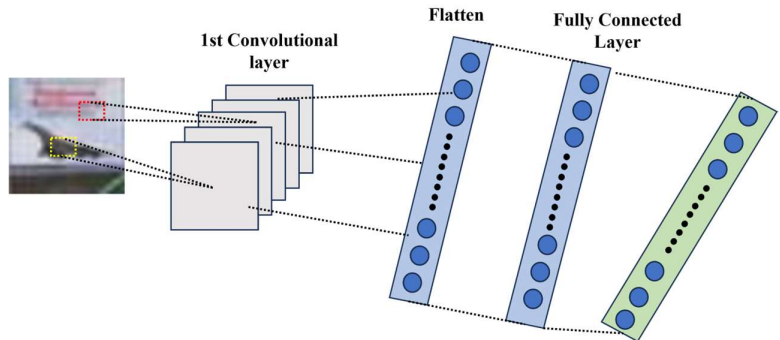

Fig. 4 The design of the convolutional layer and fully connected layer, integrated with memcapacitor crossbars, tailored for processing the MNIST dataset.

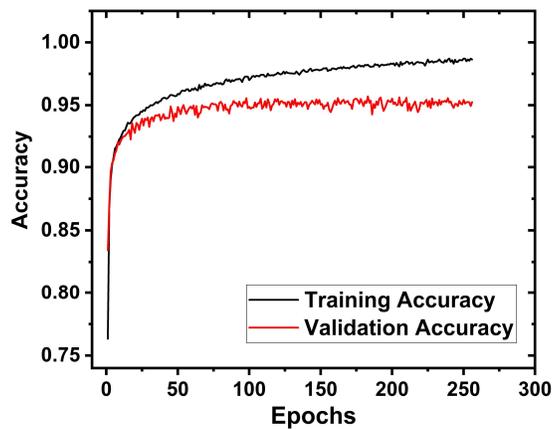

Fig. 5. MNIST training and validation accuracy per epoch of memcapacitor based VMM.

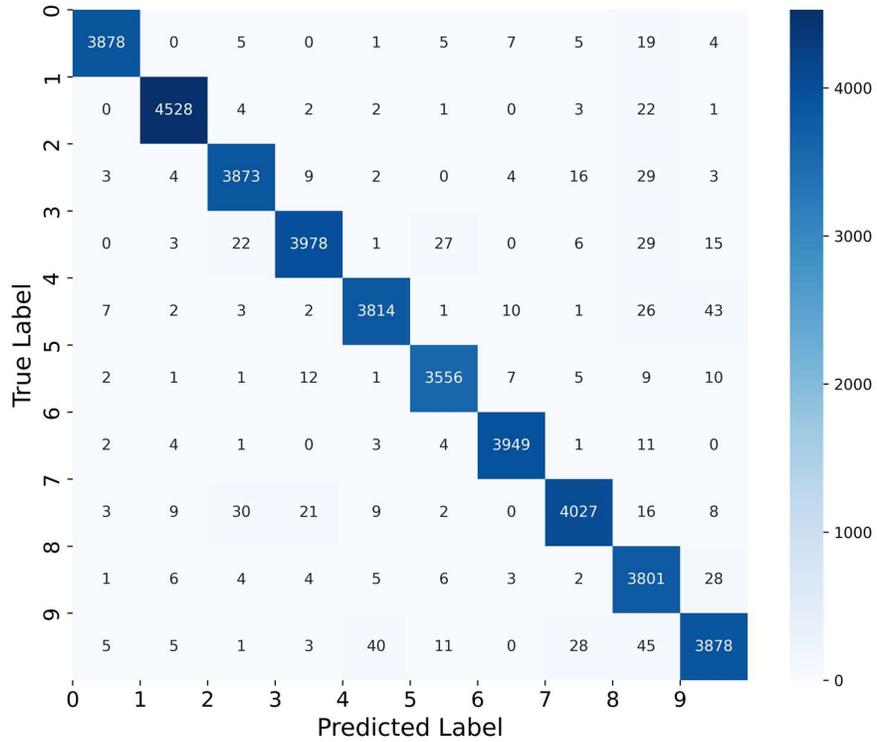

Fig. 6. Confusion Matrix of Model Predictions Against True Labels.

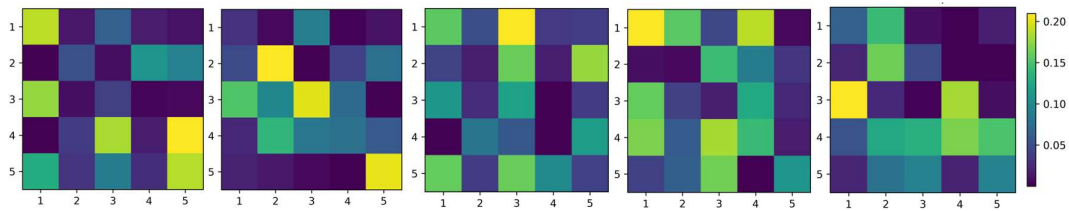

Fig. 7. Evaluating Model Robustness in Speech Recognition: Training and Validation Accuracy per Epoch.

---

**Algorithm 2 Memcapacitor based MNIST Classification**

    ***Input:***
    ***DATASET:*** *MNIST dataset*
    ***CNN model:*** *Define custom VMM model based on memcapacitor model*
    ***Model Parameters:*** *Define activations functions, fully connected layer*
    ***Parameters:*** *Training Parameters e.g. epochs, batch size*
    ***Output:,*** *Trained network, accuracy metrices, plot*
    ***Initialize*** *Normalize dataset and resize to 20 X 20*
    ***Train model:***
        *5 X 5 X 5 VMM model based on memcapacitor*
        *Flatten layer (16 X 16 to 1D)*
        *Fully Connected Layer of 64 neuron*
        *ReLU activation*
        *Fully Connected Layer of 10 neuron*
        *Softmax activation*
    ***Evaluate model such as validation accuracy***
    ***Analyze and plot simulation results***

# 5. Conclusion

In this research, we have successfully demonstrated the potential of a memcapacitor-based neural network framework for data-intensive tasks such as digit classification and image recognition. By leveraging the unique characteristics of memcapacitors, we were able to design and implement a CMOS-based circuit that not only emulates synaptic behavior but also performs vector matrix multiplication (VMM) efficiently. The proposed system was validated using the MNIST and CIFAR datasets, where it achieved impressive training accuracies of 98.4% and 94.4%, respectively, highlighting its effectiveness in neuromorphic computing applications. The results of our study indicate that memcapacitor-based systems can significantly reduce the energy demands associated with training and inference in neural networks, making them a viable solution for future AI hardware designs. The integration of memcapacitor technology within the neural network architecture has proven to enhance computational efficiency while maintaining high accuracy, setting a new benchmark for energy-efficient computing.

Our work opens the door for further exploration into memcapacitor-based hardware accelerators, particularly in the field of artificial intelligence and neuromorphic computing. Future research can build on these findings by exploring the scalability of this approach, the integration with other emerging technologies, and the development of more complex neural network architectures. Ultimately, this study underscores the potential of memcapacitor-based systems to revolutionize the design of low-power, high-performance computing devices, paving the way for more sustainable and efficient AI-driven technologies.

# Acknowledgment

This research was supported by Nanomaterial Technology Development Program through the National Research Foundation of Korea (NRF) funded by the Ministry of Science and ICT under Grant NRF-2022M3H4A1A01009658.